 \documentstyle[preprint,tighten,eqsecnum,aps]{revtex}

\preprint{THU-93/13}

\begin{document}
\draft
\preprint{}
\title{Relativistic meson spectroscopy
in momentum space
}
\author{H. Hersbach}
\address{
Instituut voor Theoretische Fysica, Universiteit Utrecht,\\
Princetonplein 5, P.O.~Box~80.006, 3508~TA Utrecht, the Netherlands}
\date{\today}
\maketitle
\begin{abstract}
In this paper a relativistic quantum theory introduced by de Groot and
\mbox{Ruijgrok} in 1975, is used as a quark model in
momentum space. The complete spectrum, with the exception of the
selfconjugate light unflavoured mesons, is calculated.
The potential used consists of an one-gluon-exchange (OGE) part and a
confining part.
For the confining part a relativistic
generalization of the linear plus constant potential was used, which is
well-defined in momentum space without introducing any singularities.
For the OGE part several potentials were investigated.
Retardations were included at all places.
Using a fitting procedure involving 52 well-established mesons,
best results were
obtained for a potential consisting of a purely vector Richardson
potential and a purely scalar confining potential.
In this way a vector confining is entirely induced by the Richardson
potential.
Reasonable results were also obtained for
a modified Richardson potential.
Most meson masses, with the exception of the $\pi$, the $K$
and the $K_0^*$, were found to be
reasonably well described by the model.
\end{abstract}

\narrowtext
\section{Introduction}
In principle the properties of mesons and baryons should be correctly
described by Quantum chromodynamics (QCD).
However, apart from some lattice gauge calculations, this is
practically impossible at the moment.
As a replacement simple quark models, in which
hadrons are viewed as bound states of constituent quarks, are quite
successful (for a review see \cite{rev}).
The simplest are the nonrelativistic (NR)
ones. The potential
used here, normally consists of a Coulomb term to
account for the perturbative
one gluon exchange (OGE), and a linear potential with possibly an
additional constant to
incorporate the nonperturbative confining. These models work very well
for the heavier charmonia and bottonia. For the lighter mesons however,
it is clear that relativistic corrections must be included. Roughly
speaking, this can be achieved in two ways. The most direct way is to
replace all NR expressions by their relativistic counterparts.
Spin dependences, like spin-spin, spin-orbit and tensor couplings are
included by hand. The second way, which, from a theoretical point of view
is more consistent,
is to start from a framework that is manifest Lorentz covariant from the
outset. The most natural representation for such a framework, like the
Bethe-Salpeter equations, is momentum space. Traditionally, however, and
also because of
the belief that it would be impossible to describe a confining potential
in momentum space, the equations are normally transformed to
configuration space.
Since a few years \cite{Vary,Heiss,gross,Norbury,maung,h2},
however, it has been realized, that
there is no obstacle to define a confining potential in momentum space,
even in the relativistic case. Therefore a growing interest has arisen to
study quark models directly in this more favourable representation.

This will also be the subject of the present paper.
The theory introduced by de Groot and Ruijgrok \cite{GR,RA,RG,HR,h1}
will be used as a model to calculate the masses of all known mesons, with the
exception of the selfconjugate light unflavoured ones. This Lorentz
covariant theory is defined via a natural generalization of the NR
Lippmann-Schwinger equation and does not require further specification in
the course of its solution. It does not start from the Bethe Salpeter
equations. The main difference is that in the intermediate states all
particles remain on their mass shell, and that total three velocity rather
than four momentum is conserved. Negative energy states are not
included. Retardation
effects are incorporated in a simple and unambiguous way. This is to be
contrasted with the Bethe-Salpeter equations,
where different three-dimensional
quasi-potential reductions lead to different retardations (see e.g.
Sec.\ (2.3) of \cite{rev}).
The theory has proven to give the correct fine structure
formulas for the hydrogen atom and positronium \cite{h1}.
In Sec. \ref{sec:theory} a brief summary of the
theory will be given.
In Sec.\ \ref{sec:pot} a number of quark-antiquark potentials
will be discussed. A modification
of the Richardson potential, to acount
for the OGE, as well as a relativistic generalization of the constant
potential is defined.
An important feature of the mesons consisting of light quarks is the
appearance of linear Regge trajectories. Their origin in the light of
the present theory is discussed in Sec.\ \ref{sec:regge}.
The numerical method used will be described in Sec.\
\ref{sec:num}, and its results
will be further discussed in Sec. \ref{sec:discus}.
The paper ends with some conclusions.

\section{Formulation of the Theory}\label{sec:theory}
In this section a summary of the theory, as introduced by de Groot and
Ruijgrok \cite{GR,RA,RG,HR}, with the modifications made in \cite{h1},
will be given
\subsection{The general framework}
A state $\alpha$ of a quark (mass $m_1$) and an antiquark (mass $m_2$),
can be characterized by $(p_1\lambda_1,p_2\lambda_2)$, where
$p_1$ and
$p_2$ are the four-momenta of the quark and antiquark, and $\lambda_1$
and $\lambda_2$ are
their helicities. Both particles are supposed to remain on
their mass shell, which means that
$p_i^0=\sqrt{|{\bf p}_i|^2+m_i^2}\equiv E_i$, $i=1,2$.
The theory is constructed in such a way that
in the interaction the three-velocity
\begin{equation}\label{veldef}
{\bf v}=\frac{{\bf p}_1+{\bf p}_2}{p_1^0+p_2^0},
\end{equation}
is conserved. This means that the quark-antiquark potential
$V_{\beta\alpha}$ for a transition from an initial state
$\alpha=(p_1\lambda_1,p_2\lambda_2)$ to a final state
$\beta=(p_1^\prime\lambda_1^\prime,p_2^\prime\lambda_2^\prime)$ contains
only non-zero elements if ${\bf v}^\prime={\bf v}$. In the center of
momentum system (cms) this velocity conservation coincides with
three-momentum conservation,
i.e., ${\bf p}_1=-{\bf p}_2\equiv {\bf p}$ and
${\bf p}_1^\prime=-{\bf p}_2^\prime\equiv {\bf p}^\prime$.
In this frame therefore the potential can be written as
\begin{displaymath}
V_{\beta\alpha}=V_{\lambda_1^\prime\lambda_2^\prime,\lambda_1\lambda_2}
({\bf p}^\prime,{\bf p}).
\end{displaymath}
In the NR case the momentum dependence of a central potential appears
in the form $|{\bf q}|^2=|{\bf p}^\prime-{\bf p}|^2$.
In the relativistic case this expression must be replaced by a covariant
one. Here the usual replacement
$|{\bf q}|^2\rightarrow -q^2$ cannot be used
because, due to the lack of four-momentum conservation, the
loss of momentum $q_1=p_1-p_1^\prime$ by the quark, will in general
differ from the gain of momentum $q_2=p_2^\prime-p_2$ by the antiquark.
Instead the following obvious and symmetrical substitution is made
\begin{equation}\label{q1q2}
|{\bf q}|^2\;\rightarrow\;-q_1\cdot q_2=
|{\bf p}^\prime-{\bf p}|^2-\tau(p^\prime,p),
\end{equation}
where the term $\tau$, defined by
\begin{equation}\label{retard}
\tau(p^\prime,p)=(E_1-E_1^\prime)(E_2^\prime-E_2),
\end{equation}
is responsible for retardation effects.
The theoretical justification for this replacement is twofold.
In the first place it can be shown \cite{h1} to be consistent with
velocity conservation. The second and more practical justification is,
that in the
case of the Coulomb potential, $\tau$
automatically generates the correct form for
the Breit interaction (see \cite{h1}).
In the equal mass case $\tau=-(E-E^\prime)^2$, which
is exactly opposite to the retardation
used by Gross and Milana \cite{gross} and
Maung, Kahana and Norbury \cite{maung}.
The difference in sign will give
the wrong sign for the Breit interaction,
which in turn will effect the
fine structure of positronium. In \cite{h2} it was shown
that Eq.\ (\ref{retard}) gives rise to the correct
positronium fine structure
formula.

The relativistic wave equation from which
the mass $M$ of the meson is
to be solved is in the cms given by ($\hbar=c=1$)
\begin{eqnarray}\label{waveequation}
[E_1+E_2-M]\Psi_{\lambda_1\lambda_2}({\bf p})+&&\\
\sum_{\lambda_1^\prime\lambda_2^\prime}
\int W_{\lambda_1^\prime\lambda_2^\prime,\lambda_1\lambda_2}
({\bf p}^\prime,{\bf p})\Psi_{\lambda_1^\prime\lambda_2^\prime}({\bf
p}^\prime)
\left[\frac{m_1m_2}{E_1^\prime E_2^\prime}\right]d{\bf p}^\prime&=&0,
\nonumber
\end{eqnarray}
where the wave function $\Psi_{\lambda_1\lambda_2}({\bf p})$
is normalized as
\begin{equation}\label{fnorm}
\sum_{\lambda_1\lambda_2}\int |\Psi_{\lambda_1\lambda_2}({\bf p})|^2
\left[\frac{m_1m_2}{E_1E_2}\right]d{\bf p}=1
\end{equation}
and $V=4m_1m_2W$. The quantity $W$ is introduced for convenience, because
it reduces in the NR limit to the NR potential. In this limit Eq.\
(\ref{waveequation}) reduces to the NR Schr\"odinger equation in
momentum space.

\subsection{Decomposition}
The interaction $W$ used in this paper can be decomposed
into a vector part $V_V$ and a scalar part $V_S$, which is in the cms
given by
\begin{eqnarray}\label{vvs}
W({\bf p}^\prime,{\bf p})&=&
\overline{u}_{\lambda_1^\prime}({\bf p}_1^\prime)
\overline{v}_{\lambda_2}({\bf p}_2)
\left[\gamma_\mu^{(1)}\cdot\gamma^{(2)\mu}
V_V({\bf p}^\prime,{\bf p}) \right.\nonumber\\
& &\left.+\openone^{(1)}\openone^{(2)} V_S({\bf p}^\prime,{\bf p})\right]
v_{\lambda_2^\prime}
({\bf p}_2^\prime)
u_{\lambda_1}
({\bf p}_1).
\end{eqnarray}
Here the Dirac spinors $u$ and $v$ for the quark resp. antiquark, are
defined by
\begin{eqnarray}\label{uvdef}
u_\lambda({\bf p})&=& N \left[\begin{array}{c}1\\2\lambda b\end{array}
\right]\chi(\lambda,\frac{{\bf p}}{p}),\nonumber\\
v_\lambda({\bf p})&=& N \left[\begin{array}{c}-2\lambda b\\1\end{array}
\right](-i)\sigma_2\chi^{*}(\lambda,\frac{{\bf p}}{p}),\nonumber
\end{eqnarray}
with $N=\sqrt{(E+m)/(2m)}$, $b=p/(E+m)$, and $\chi(\lambda,{\bf p}/p)$ the
helicity spinor with helicity $\lambda$.
For the two-particle helicity states we use the conventions introduced by
Jacob and Wick \cite{jacob}.
Potential (\ref{vvs}) partially decouples with respect to the states
$|p;JM;\lambda_1\lambda_2\rangle$, which are defined by Eq.\ (18) of
\cite{jacob}, giving
\begin{eqnarray}\label{Jdec}
\langle p^\prime;J^\prime M^\prime;\lambda_1^\prime\lambda_2^\prime|W|
p;JM;\lambda_1\lambda_2\rangle&=&\nonumber\\
\delta_{JJ^\prime}\delta_{MM^\prime}
\langle\lambda_1^\prime\lambda_2^\prime|W^J(p^\prime,p)|
\lambda_1\lambda_2\rangle.&&
\end{eqnarray}
Because of conservation of parity, $W$ further decomposes into two $2\times
2$ submatrices, each having a definite parity. The subspace spanned by
\begin{eqnarray}\label{triplet}
|t_1\rangle&=&\frac{1}{\sqrt 2}\left[|\frac{1}{2},\frac{1}{2}\rangle+
|-\frac{1}{2},-\frac{1}{2}\rangle\right],\\
|t_2\rangle&=&\frac{1}{\sqrt 2}\left[|\frac{1}{2},-\frac{1}{2}\rangle+
|-\frac{1}{2},\frac{1}{2}\rangle\right],\nonumber
\end{eqnarray}
has parity $(-1)^{J+1}$. It contains the triplet $J=l\pm1$ states.
The complementary subspace, spanned by
\begin{eqnarray}\label{singlet}
|s_1\rangle&=&\frac{1}{\sqrt 2}\left[|\frac{1}{2},\frac{1}{2}\rangle-
|-\frac{1}{2},-\frac{1}{2}\rangle\right],\\
|s_2\rangle&=&\frac{1}{\sqrt 2}\left[|\frac{1}{2},-\frac{1}{2}\rangle-
|-\frac{1}{2},\frac{1}{2}\rangle\right],\nonumber
\end{eqnarray}
has parity $(-1)^J$ and contains the $J=l$ singlet and triplet states.
Only in the equal mass case this subspace further splits into two
$1\times1$ subspaces.
Let
\begin{equation}\label{vdec}
V^{nJ}_{ij}=p^\prime p\frac{m_1m_2}{\sqrt{E_1^\prime E_2^\prime E_1E_2}}
\langle n_i|W^J(p^\prime,p)|n_j\rangle,\;\;n=s,t,
\end{equation}
then eigenvalue Eq.\ (\ref{waveequation}) can be cast in the form
(suppressing the quantumnumbers $J$, $M$ and $s$ or $t$)
\begin{equation}\label{decompeq}
\left[E_1+E_2-M\right]f_i(p)+\sum_j\int_0^\infty V_{ij}(p^\prime,p)
f_j(p^\prime)dp^\prime=0.
\end{equation}
In appendix \ref{app:partwave} explicit formula's for
$V_{ij}^{nJ}=(V_V)_{ij}^{nJ}+(V_S)_{ij}^{nJ}$
are given.
The reduced wave function $f$ is normalized to
\begin{equation}\label{normred}
\sum_i\int_0^\infty|f_i(p)|^2dp=1.
\end{equation}

\section{The quark-antiquark potential}\label{sec:pot}
The quark-antiquark potential must contain an one-gluon exchange (OGE)
to account for the short range, and a confining part for the
long range interaction. It is generally believed that $V_{\rm OGE}$
should have a vector Lorentz
structure, while about the confining part $V_{\rm CON}$ there is no
consensus. Some believe that it must have a purely scalar structure,
while others admit a mixture between scalar and vector coupling.
We will adopt this last point of view.
The potential can therefore
be written in the form (see Eq.\ (\ref{vvs}))
\begin{eqnarray}\label{vqq}
V_V&=&V_{\rm OGE}+\epsilon V_{\rm CON},\nonumber\\
V_S&=&(1-\epsilon) V_{\rm CON},
\end{eqnarray}
where $\epsilon$ is the scalar-vector mixing of the confining potential.

For $V_{\rm CON}$
the relativistic generalization of the linear potential, as described in
\cite{h2}, plus a constant potential (to be defined below) was used.
This generalization is defined in a formal way and does not introduce any
singularities.
For the OGE two different potentials were used:
the Richardson potential \cite{rich} and
a modified version of this potential,
both containing a running coupling constant (see Sec.\
\ref{sec:runcop}).
The Richardson potential contains a linear part by itself.
Therefore in this case
(from now on denoted by I), $\epsilon=0$ was chosen, so that
the confining in the
vector direction is completely determined by the Richardson
potential. The modified Richardson potential has no linear part.
Therefore in this case (denoted by II) a nonzero $\epsilon$ was admitted.

In all these potentials the
NR momentum transfer $|{\bf q}|^2$ was replaced by
$-q_1\cdot q_2$ (see Eq.\ (\ref{q1q2})) and in this way
retardation effects were included everywhere. For notational convenience,
the quantity
\begin{equation}\label{Qdef}
Q\equiv \sqrt{(-q_1\cdot q_2)}
\end{equation}
is introduced. In the NR limit it reduces to ${|\bf q}|$.

\subsection{The one-gluon-exchange:\protect\\running coupling constant}
\label{sec:runcop}
The renormalization scheme of perturbative QCD says that, for large
momentum transfer, the
running coupling constant $\alpha_s$
as it occurs in the one gluon exchange
(the factor $\frac{4}{3}$ arises from color averaging)
\begin{equation}\label{alphas}
V_{\rm OGE}= -\frac{4}{3}\frac{\alpha_s(Q^2)}{2\pi^2Q^2}
\end{equation}
is given by (see also Eq.\ (B.2) of \cite{data})
\begin{equation}\label{alphaqcd}
\alpha_s(Q^2)=\frac{a_n}{X_n}\left[1-b_n\frac{\log(X_n)}{X_n}+{\cal O}
\left(\frac{\log^2(X_n)}{X_n^2}\right)\right].
\end{equation}
Here
\begin{displaymath}
a_n=\frac{12\pi}{(33-2n)},\hspace{0.5cm}
b_n=\frac{6(153-19n)}{(33-2n)^2},
\end{displaymath}
\begin{displaymath}
X_n=\log\left[Q/\Lambda^{(n)}_{\overline{\rm MS}}\right]^2,
\end{displaymath}
and $n$ is the number of quarks with a mass smaller than $Q$.
The subscript $\overline{MS}$ denotes that the renormalization is
performed according to the modified minimal subtraction scheme.
The connection between the different $\Lambda^{(n)}_{\overline{\rm MS}}$'s
is given by Eq.\ (B.4) of \cite{data}.
The typical momentum transfer within a meson is on the order of one GeV,
so in this region $n=3$.
Therefore, Eq.\ (\ref{alphaqcd}) with $n=3$ is in many cases used as an
approximation for all
large momentum transfers. In addition the $b_3$ term is almost always
neglected. But this term is not small at all: in the $Q$-region from 1 to
5 GeV its contribution is about 25\%. Even for very high momentum
transfers its contribution is substantial $\sim {\rm 15}\%$ for $Q={\rm 50\;
GeV}$.
However, it appears that when $\Lambda_{\overline{MS}}^{(5)}$
rather than $\Lambda_{\overline{MS}}^{(3)}$ is used, a fairly
good approximation of Eq.\ (\ref{alphaqcd})
for large $Q$ is obtained by putting
\begin{equation}\label{alsapp}
\alpha_s\approx\frac{a_3}{X_5}
\end{equation}
(see the curve ``standard
approximation'' of Fig. \ref{fig1}).
For $Q={\rm 5\;GeV}$ the deviation from Eq.\ (\ref{alphaqcd})
is $\sim {\rm 7}\%$,
and for $Q\sim {\rm 50\;GeV}$ there is no detectable
difference. Also for smaller $Q$
the agreement is better, but of course in this region the validity of
Eq.\ (\ref{alphaqcd}) is doubtful. Nevertheless we think that
these considerations show
that there is no theoretical necessity
to stick to the value of $a_3$ in Eq.\
(\ref{alsapp}): a small deviation from it also results in a good
running coupling constant for large $Q$.

For small positive $Q$-values Eqs.\ (\ref{alphaqcd}) and (\ref{alsapp})
diverge.
To remedy this, Richardson \cite{rich}
proposed a potential in which the
divergence is shifted to the origin by making the replacement
$Q^2\rightarrow Q^2+\Lambda^2$ in Eq.\ (\ref{alsapp}):
\begin{eqnarray}\label{richard}
V_R(Q^2)&=&-\frac{\alpha_0}{2\pi^2Q^2\log[1+\frac{Q^2}{\Lambda^2}]}\\
&=& -\frac{\alpha_0\Lambda^2}{2\pi^2Q^4}
-\frac{\alpha_0}{4\pi^2Q^2}+\ldots
\;\mbox{for}\;Q\rightarrow 0.\nonumber
\end{eqnarray}
The color factor $\frac{4}{3}$ is absorbed in $\alpha_0$.
In Fig.\ \ref{fig1} the running coupling constant,
defined via Eq.\ (\ref{alphas}) with
$V_{\rm CON}=V_{\rm R}$ is
compared to the QCD formula
for $\alpha_0=\frac{4}{3}a_3=16\pi/27={\rm 1.862}$.
The singularity for $Q=0$ results from a linear term in the
potential with string tension
$\frac{1}{2}\alpha_0\Lambda^2$ (see \cite{h2} or Sec.\ \ref{sec:lin}).
When the singularity is subtracted, the running coupling
constant saturates to the value $\frac{1}{2}a_3=0.698$.
{}From Fig.\ \ref{fig1} it is seen that for
momentum transfers starting from 2 GeV, a much better approximation to
Eq.\ (\ref{alphaqcd}) is obtained, if $\alpha_0$ is slightly decreased.
A value of
$\alpha_0={\rm 1.750}$
turns out to be a very good choice.

A different way to remove the singularities is to also make the
replacement $Q^2\rightarrow Q^2+\Lambda^2$ in $1/Q^2$ itself. This
results into
\begin{eqnarray}\label{vh}
V_M(Q^2)&=&-\frac{\alpha_0}{2\pi^2[Q^2+\Lambda^2]
\log[1+\frac{Q^2}{\Lambda^2}]}
\\&=&-\frac{\alpha_0}{2\pi^2Q^2}+\ldots\;\mbox{for}\;
Q\rightarrow 0.\nonumber
\end{eqnarray}
This modified Richardson potential $V_M$
does not contain a linear part. The
coupling constant saturates to a value $a_3$.
The running coupling constant defined via this potential
is given in Fig.\ \ref{fig1} for $\alpha_0=\frac{4}{3}a_3$. {}From
this figure it is seen that this choice for $\alpha_0$ gives a good
representation of the QCD formula for moderate $Q$ values.

The spinless partial waves $W_R$ and $W_M$ of $V_R$
and $V_M$, defined by Eq.\
(\ref{vnospin}), are given in appendix \ref{app:rich}.

\subsection{The confining:\protect\\ ``linear + constant'' potential}
\label{sec:lin}
For the confinement a relativistic generalization of a linear plus
constant potential was used:
\begin{displaymath}
V_{\rm CON}=V_{\rm LIN}+V_{\rm C}.
\end{displaymath}
As was already mentioned in the introduction, it was
for a long time believed that
a linear potential could not correctly be described in momentum space.
A naive consideration
shows that it behaves like $-1/Q^4$, which results in
an ill-defined bound state equation. A few years ago
it was shown \cite{Vary,Heiss,Norbury}
that this singularity for the NR case is only
apparent. For the relativistic case different methods were employed
\cite{gross,maung,h2} to solve this problem.
The one used in this paper is defined by \cite{h2}
\begin{equation}\label{vlindef}
V_{\rm LIN}=\lim_{\eta\downarrow 0}\frac{\partial^2}{\partial\eta^2}
\frac{\lambda}{2\pi^2}\left[\frac{1}{Q^2+\eta^2}\right],
\end{equation}
where the color factor $\frac{4}{3}$ is absorbed in the string tension
$\lambda$.
In \cite{h2} it was shown that the limit exists in a distributional sense.
The result was that the integral in Eq.\ (\ref{decompeq}) is replaced by
\begin{equation}\label{principal}
-\hspace{-0.4cm}\int_0^\infty\left[ V_{ij}^{(nJ)}(p^\prime,p)f_j(p^\prime)
+\frac{4p^2 C_{ij}(p)}{(p^{\prime2}-p^2)^2} f_j(p)\right]dp^\prime.
\end{equation}
Here $V^{nJ}_{ij}$, $n=s,t$, is the naive pointwise limit obtained from
Eq.\ (\ref{vlindef}). The $1/(p^\prime-p)^2$ singularity is removed by the
quantity
\begin{equation}\label{Cdef}
C_{ij}(p)=\lim_{p^\prime\rightarrow p}\left[-[p^\prime-p]^2
V_{ij}^{(nJ)}(p^\prime,p)\right].
\end{equation}
The resulting $1/(p^\prime-p)$ singularity
is handled by the principal value
integral (denoted by $-\hspace{-0.3cm}\int$).
It was shown that this subtraction is not just a trick to avoid
singularities, but is really generated by Eq.\ (\ref{vlindef}).

For a confining potential that consists of a mixture of
a scalar and vector Lorentz
structure (see Eq.\ (\ref{vqq})), the pointwise limits of the spinless
partial waves (see Eq.\ (\ref{vnospin})) are given by $W^l_V=\epsilon W^l_L$
and $W^l_S=(1-\epsilon)W^l_L$, where
\begin{equation}\label{Wnaive}
W^l_L(p^\prime,p)=\frac{\lambda}{\pi}R(p^\prime,p)\frac{Q_l^\prime(z_0)}
{p^\prime p}
\end{equation}
and also $R(p^\prime,p)$ is given in Appendix\ \ref{app:partwave}.
Here $z_0$ is defined by Eq.\ (\ref{z0}) and $Q_l$ is the Legendre function
of the second kind of order $l$.
The $1/(p^\prime-p)^2$ singularity of $W_L^l$ is determined by
\begin{displaymath}
-\frac{\lambda}{\pi}\frac{R(p,p)}{(p^\prime-p)^2-\tau(p^\prime,p)}
\end{displaymath}
The retardation defined by Eq.\ (\ref{retard}) behaves around
$p^\prime\approx p$ like
\begin{displaymath}
\tau(p^\prime,p)=-\frac{p^2}{E_1E_2}\left(p^\prime-p\right)^2+\ldots
\end{displaymath}
and therefore contributes to the singularity. It follows that
$C_{ij}$ is given by
\begin{equation}\label{Clin}
C_{ij}(p)=\frac{\lambda}{\pi}\left[\epsilon+(1-\epsilon)\frac{m_1m_2}
{p_1\cdot p_2}\right]\delta_{ij},
\end{equation}
where $p_1\cdot p_2=E_1E_2+|{\bf p}|^2$ is the dotproduct between the
four vectors $p_1$ and $p_2$.
Note that $C_{ij}$ does not depend on $J$
and the parity $s$ or $t$. In addition it is a manifest Lorentz covariant
object.

When the interaction does not contain a linear part, the integral
(\ref{principal}) coincides with the integral within Eq.\ (\ref{decompeq}).
This is so because then $C_{ij}=0$ and there is no $1/(p^\prime-p)$
singularity, which means
that the principal value coincides with an ordinary
integral. Therefore replacement (\ref{principal}) in combination with Eq.\
(\ref{Cdef}) can be applied to
the entire interaction. In this way a nonzero
value of $C$ automatically
indicates the presence of a linear term. For the
Richardson potential (\ref{richard}) with a purely vector character this
results in
\begin{equation}\label{Crich}
(C_R)_{ij}=\frac{\alpha_0\Lambda^2}{2\pi}\delta_{ij},
\end{equation}
which indicates a linear term with string tension
$\frac{1}{2}\alpha_0\Lambda^2$.

In analogy with the linear potential the constant potential $V_C$ can
also be
defined via the Yukawa potential. In the NR case one has in configuration
space
\begin{displaymath}
V_C(r)=C=\lim_{\eta\downarrow 0}\frac{\partial}{\partial\eta}
\left[-\frac{Ce^{-\eta r}}{r}\right].
\end{displaymath}
Therefore in momentum space an obvious relativistic generalization is
\begin{equation}\label{VCdef}
V_C(Q^2)=\lim_{\eta\downarrow 0}\frac{\partial}{\partial\eta}
\left[-\frac{C}{2\pi^2(Q^2+\eta^2)}\right].
\end{equation}
Note that this expression also includes retardations, which are hidden in
$Q^2$ (see Eq.\ (\ref{Qdef})). Definition (\ref{VCdef}) has to be included
in the integral of Eq.\ (\ref{decompeq}) before the limit is taken.
The spinless partial wave $W_\eta^l$ of this constant potential is given by
\begin{displaymath}
W_\eta^l(p^\prime,p)=-\frac{CR(p^\prime,p)}{\pi}
\frac{\eta}{p^\prime p}Q_l^\prime\left[
z_0+\frac{\eta^2}{2p^\prime p}\right].
\end{displaymath}
The only term that survives the limit $\eta\downarrow 0$ is
\begin{displaymath}
\frac{CR}{\pi}\frac{\eta}{(p^\prime-p)^2-\tau+\eta^2}\rightarrow
CR\left[\frac{E_1E_2}{p_1\cdot p_2}\right]^{\frac{1}{2}}
\delta(p^\prime-p)+\cdots
\end{displaymath}
For a confining potential that has both a scalar and
vector part (see Eq.\ (\ref{vvs})) this results in
\begin{equation}\label{VC}
(V_C)^{nJ}_{ij}=C\left[\frac{p_1\cdot p_2}{E_1E_2}\right]^\frac{1}{2}
\left(\epsilon+(1-\epsilon)\frac{m_1m_2}{p_1\cdot p_2}\right)
\delta_{ij}\delta(p^\prime-p).
\end{equation}

\section{Linear Regge trajectories}\label{sec:regge}
The mesons which consist of the light $u$, $d$ and $s$ quarks only, are
found to lie on so-called linear Regge trajectories. This means that
there are several groups of mesons, for which the mass squared for each
meson within such a group, is proportional to its angular momentum $J$,
i.e. $M_J^2\approx\beta J+C$. The constant $C$ depends on the group, the
Regge slope $\beta$ however is about the same for all groups. Its
experimental value is $\beta\approx1.2\;({\rm GeV})^2$. For
mesons containing a heavy $c$ or $b$ quark, such trajectories have not
been observed. This makes it plausible that linear trajectories are
induced by relativistic effects. In fact, it is known
\cite{rev,bas,Martin},
that the Schr\"odinger equation with
ultrarelativistic (UR) kinematics (i.e. $2p$ instead of $p^2/(2\mu)$) for
a linear potential, does indeed give rise to linear trajectories, while the
(NR) Schr\"odinger equation does not. The slope $\beta$ solely depends on
the string tension $\lambda$, namely $\beta=8\lambda$.

For the present case a similar effect is observed. It numerically appears
that the (UR) limit (i.e. $m_1,m_2\rightarrow 0$)
of bound state equation (\ref{decompeq}) also leads to linear
trajectories,with a group independent slope $\beta$. This slope however,
depends on the vector part $\lambda_V$ of the string
tension
only. In addition, the dependence is a factor $\sqrt 2$ larger than for
the relativized Schr\"odinger equation, namely
\begin{equation}\beta\approx(8\sqrt 2)\lambda_V.\label{Regge}\end{equation}
As can be deduced from Appendix \ref{app:partwave}, the off-diagonal
elements $V_{12}$ and $V_{21}$ of both a vector and a scalar potential
vanish in the (UR) limit. Therefore Eq.(\ref{decompeq}) further decouples
into two single equations.
For the pure vector case, it reduces for the $V_{11}^{tJ}$ channel to
\begin{equation}\label{ultra}
\left[2p-M\right] f_J(p)+-\hspace{-0.4cm}\int_0^\infty\hspace{-0.1cm}
\left[V^J(p^\prime,p)
f_J(p^\prime)+\frac{\lambda}{\pi}f_J(p)\right]\!dp^\prime=0,
\end{equation}
with
\begin{equation}\label{wijultra}
V^J(p^\prime,p)=\frac{2\lambda}{\pi p^\prime p}Q_J^\prime \left[
\frac{p^{\prime 2}+p^2+(p^\prime -p)^2}{2p^\prime p}\right].
\end{equation}
This equation was solved numerically using the method described in
Section~\ref{sec:num}.
The calculated masses (in units of $\sqrt\lambda$)
of the lowest states for each $J$
are presented in Table~\ref{regge}.
The Schr\"odinger equation with (UR) dynamics is in momentum space also
given by Eq.\ (\ref{ultra}), but now with
\begin{equation}\label{zijultra}
V^J(p^\prime,p)=\frac{\lambda}{\pi p^\prime p} Q_J^\prime
\left[ \frac{p^{\prime 2}+p^2}{2p^\prime p}\right].
\end{equation}
The corresponding calculated masses are also listed in Table~\ref{regge}.
They all agree with the calculations performed by Basdevant and Boukraa
(see Table I of \cite{bas}).

In principle the trajectories are expected to be linear only for large
values of $J$.
However, as Table~\ref{regge} shows, the convergence is very fast.
It was found that also for moderate masses Eq.\ (\ref{decompeq}) leads to
Regge trajectories with the same relation (\ref{Regge}) between $\beta$ and
$\lambda$. The convergence, however, is then slower.
When in addition a OGE term and a constant are added to the potential,
relation (\ref{Regge}) is affected. The change, however, is not very
large. Therefore
it can be concluded that, in order to obtain reasonable
Regge slopes, the string tension
in the vector direction $\lambda_V$ should be
around
$\lambda_V\approx {\rm 0.1}\;{\rm GeV}^2$.

\section{Numerical method}\label{sec:num}
The present model embraces
eigenvalue equation (\ref{waveequation}) in combination with
a quark-antiquark potential $W$ consisting of an
one-gluon-exchange part $V_{\rm OGE}$ and a confining part $V_{\rm CON}$.
The way in which $V_{\rm OGE}$ and $V_{\rm CON}$ enter
in eigenvalue equation
(\ref{waveequation}) is given by Eqs.\ (\ref{vvs}) and (\ref{vqq}).
The OGE potential is determined by two parameters $\alpha_0$ and
$\Lambda$ and the confining potential is determined by a string tension
$\lambda$ and a constant $C$. Furthermore a parameter $\epsilon$ can be
introduced to give the confining potential a mixed scalar-vector
character.

The numerical solution of the model can be divided into two parts.
The first one concerns the calculation of the masses of the mesons, from
eigenvalue equation (\ref{waveequation}),
given all parameters of the potential under consideration, and the quark
masses. The second part is the fitting to the experimental data. The
eigenvalue equation was solved by
expanding the wave function into cubic  Hermite
splines (see \cite{splines}). The integration region $p\in[0,\infty)$ was
projected onto the finite interval $x\in[0,1]$ by $x=(p-p_0)/(p+p_0)$,
where $p_0$ was chosen in the physical region. On this interval
$N$ equidistant spline intervals were chosen on which $2N$
spline functions were defined.
The matrix elements of the resulting eigenvalue
equation for the expansion coefficients only involved single integrations
of the potential times a spline function. This is a major advantage of
the spline method compared to the more conventional expansion
techniques, where the evaluation of
matrix elements involves two-dimensional integrals.
The integration was performed using Gauss-Legendre quadratures. In the
case where the singular point $p^\prime=p$ was inside the region of
integration, special
care had to be taken. In the first place, an even number of
abscissas centered around $p^\prime=p$ was used. In that way the
principal value, which occurs for the confining potential,
is automatically taken care of \cite{Norbury,sloan}. Secondly, the
logarithmic singularity $\sim\log(|p^\prime-p|)$, which is induced by both
the Coulomb and the confining potential, was separately handled by means
of Gaussian quadratures based on a logarithmic weigth function (see e.g.
Table 25.7 of \cite{abra}). Another important advantage of using Hermite
splines, is their small nonzero domains. Therefore on each spline
interval only a few of these
splines (four for
the Legendre and three for the logarithmic quadrature) were needed to
obtain high accuracies. The matrix equation was solved using standard
techniques \cite{numc}, giving the meson masses $M_i$.

The choice of the projection parameter $p_0$ and the number of intervals $N$,
depended on the specific meson. For instance, the typical momentum
transfer for the $\Upsilon$ mesons ($b\overline{b}$) is about 1 GeV, so
$p_0={\rm 1}\;{\rm GeV}$. The masses of these mesons are all known to a high
precision and they are all radial levels of the $J^{PC}=1^{--}$ channel. The
$\Upsilon_{v}$ is the tenth radial state ($n=10$). Therefore 20 spline
intervals were needed to guarantee accurate results. The $K_2^*$
however, is the only known
$J^{P}=2^+$ strange meson, apart from the unconfirmed $K_2^{*\prime}$.
Therefore $N={\rm 8}$ was sufficient to
obtain reliable results. $p_0={\rm 0.5}\;{\rm GeV}$ was
a proper choice for this
meson.

The second part of the problem was to get a good fit to the experimental
data. For this purpose the merit function
\begin{equation}\label{chi2}
\chi^2(a_1,.,a_n)=\sum_i\left[\frac{M^{\rm the}_i(a_1,.,a_n)-M^{\rm exp}_i
}{\sigma_i}\right]^2
\end{equation}
has to be optimized with respect to the parameters $a_1,.,a_n$.
Here $i$ labels the mesons, $M^{\rm exp}_i$ and
$M^{\rm the}_i$ denote their experimental and calculated
masses, and $\sigma_i$ their weights. A nonlinear regression method,
based on the Levenberg-Marquardt algorithm was used to perform the fits
(see section 15.5 of
\cite{numc}). This method requires as input the explicit knowledge of the
derivatives of the calculated masses with respect to the fitparameters.
For the present complex situation this information is not known.
It is only known that the derivatives of a meson mass with respect to
quark masses it does not contain, is equal to zero.
Therefore the derivatives where approximated in the
least time consuming way by the following expression:
\begin{equation}\label{crude}
\frac{\partial M^{\rm the}_i}{\partial a_j}\approx \frac{
 M_i^{\rm the}(a_1,.,a_j+\Delta,.,a_n)-
M_i^{\rm the}(a_1,.,a_n)}{\Delta}.
\end{equation}
In this manner all required information, e.g. $M^{\rm the}_i$ and
$\partial M^{\rm the}_i/\partial a_j$, is obtained by calculating all
meson masses $(n+1)$ times. A more sophisticated method would
considerably increase this number. Approximation (\ref{crude}) turned out
to be very effective: starting with a physically sensible set of
parameters, after four or five steps convergence to an
optimum was reached. The value of the parameter $\Delta$ appeared to be
of minor importance. $\Delta={\rm 0.04}$ (dimensionless, in GeV, or
${\rm GeV}^2$, depending on the dimension of $a_j$)
was found to be a good choice.

All mesons regarded to be established by the 1992 Particle Data Group
\cite{data} (in Table\ \ref{meson_spectrum} indicated by a
``$\bullet$'')
were used in the fit, with the exception of the
selfconjugate (i.e. Isospin 0) light unflavoured ones. For a fair
description of these mesons, an annihilation interaction from initial
$q\overline{q}$ states to final $q^\prime\overline{q}^\prime$ states
should be included.
Also the charmed strange $D^*_s$ and $D_{sJ}$ were excluded,
because of the uncertainty of their quantum numbers.
Furthermore the up and down quark were
considered to be of equal mass. In addition the electromagnetic
interactions are completely neglected. Therefore the $\pi^0$ and
$\pi^\pm$, the $K^0$ and $K^\pm$, and so on will be degenerate in this
picture. Because of the indistinguishability of the $u$ and $d$ quark, from
now on such a quark will be denoted by ``$u\!/\!d$''.
This accumulates  to a total of 52 mesons: 11 light unflavoured
($u\!/\!d\overline{u\!/\!d}$), 11 strange
($s\overline{u\!/\!d}$), 4 charmed ($c\overline{u\!/\!d}$),
2 charmed strange
($c\overline{s}$), 10 charmonia ($c\overline{c}$), 2 bottom
($b\overline{u\!/\!d}$) and 12 bottonia ($b\overline{b}$).

For the weight $\sigma_i$ the maximum of the uncertainty
$dM^{\rm exp}_i$ of the measured mass and the predictive power of the
model, was taken.
It is difficult to give an estimate for this predictive power. In the
first place quark models have a phenomenological nature; there is no
direct link with QCD. In the second place,
the mesons are not stable particles, but in fact resonances.
The  decay mechanisms, which are not incorporated in this
paper, could considerably effect the position of the calculated masses.
This especially applies for the mesons that have decay widths of a
few hundred MeV.
To account for all of this, a grid size
$S={\rm 20}\;{\rm MeV}$ was introduced to give a minimum to $\sigma$.
Only for bottonium ($b\overline{b}$), a grid size of 10 MeV
was used, because in this system relativistic effects are less important
and most states have narrow widths.
Summarizing, the weights were determined by the following formula
\begin{equation}\label{sigma}
\sigma_i={\rm Max}\left[dM_i^{\rm exp},S\right].
\end{equation}
A few exceptions to this rule were made. The pion $\pi$ and the kaon $K$
are the ground states of the
$u\!/\!d\overline{u\!/\!d}$ resp
$s\overline{u\!/\!d}$ mesons.
It is commonly believed, that, in order
to give a fair description of these particles the mechanism of
chiral symmetry breaking should be included in the
model. It appeared that also the $K_0^*$ mass was badly described by the
model. This state, however,
has a large decay width of $\sim {\rm 300}\,{\rm MeV}$.
Therefore $\sigma_{\pi}={\rm 0.4}\;{\rm GeV}$ and
$\sigma_{K}=\sigma_{K_0^*}={\rm 0.2}\;{\rm GeV}$
were chosen, so that
these states get an insignificant weight in the fit.
Another point are the $\rho^\prime$ and $\rho^{\prime\prime}$.
These states also have large decay widths ($\sim {\rm 300}\;{\rm MeV}$).
It appeared that best results were obtained if each state was regarded to
be composed of two neighboring resonances. In Sec.\ \ref{sec:discus}
his point will be discussed in more detail.

To decrease the computation time first a rough fit was made by taking
only half of the spline intervals $N$ needed to obtain the desired
accuracies. The resulting fitparameters were then used as the starting
point for a full accuracy fit.  The typical computation time for
a complete rough fit for all 52 mesons was 30 min on a Sparc 2 workstation,
while the fine tuning fit took about one hour.

As was already mentioned,
two different types of potentials were examined.
In case I the Richardson potential $V_{\rm R}$ was taken to account for
the OGE and for the confinement in the vector direction. $V_{\rm CON}$ has a
purely scalar character ($\epsilon=0$). For $\alpha_0$ both the ``QCD''
value $16\pi/27$ and the value 1.75, which gives a better agreement
with the QCD formula (\ref{alphaqcd}), were taken. Both choices ended in
comparable fits ($\chi^2\approx {\rm 260}$). The resulting parameters
for $\alpha_0=1.75$ (denoted by Ia) are given in Table\ \ref{fitresult}
and the calculated meson spectrum in Table\ \ref{meson_spectrum}.
Also the case in which naither $\alpha_0$ nor $\epsilon$ was fixed was
regarded. The regression method led to very small values for $\epsilon$
and the string tension $\lambda$. Therefore a fit, denoted by Ib,
was made where these two
parameters were put equal to zero, and where $\alpha_0$ was varied.
This resulted in a somewhat better fit (Ib) with $\chi^2={\rm 250}$
(see also Tables
\ref{fitresult} and \ref{meson_spectrum}).
In both cases seven parameters, three to model the potential and four
quark masses, were fit.
Finally, the case was considered in which the linear term of
$V_{\rm R}$ was subtracted.
To get a confining potential in the vector direction,
the mixing $\epsilon$ was also varied.
The results did not improve, however.

In case II the modified Richardson potential $V_{\rm M}$
in combination with a
mixed scalar-vector $V_{\rm CON}$ was taken.
The value $\alpha_0=16\pi/27$ was the only parameter held fixed, so that
eight parameters were varied. In spite of the extra parameter,
the resulting fit (II), see Tables \ref{fitresult} and \ref{meson_spectrum},
is worse than the fits found for case I and gave $\chi^2={\rm 322}$.

\section{Discussion}\label{sec:discus}
The meson spectrum calculated for parameter sets Ia, Ib and II is given
in Table\ \ref{meson_spectrum}. Also the mesons that were not involved in
the fitting procedure (the ones without a $\sigma$) were calculated.
It is seen that most of these unconfirmed mesons
(see \cite{data}), are reasonably described by the model. Many states are
a mixture between two ${}^{2s+1}L_J$ waves. Only in the NR limit these
waves decouple because only then the angular momentum $l$ is a good
quantum number. For each state the most dominant wave is underlined. Most
distributions are like 99\% vs. 1\%, which supports the statement that
$l$ is almost a good quantum number.

A few years ago (for a review, see page VII of \cite{data}), the
$\rho$(1450) and $\rho$(1700) were recognized as being a splitting in the
formerly known $\rho$(1600) resonance. It could be interpretated as the
fine structure
splitting between the $n=2$, dominantly $S$, and the $n=3$,
dominantly $D$, states. The splitting however is rather big ($\sim$ 250
MeV). For the present model this interpretation was found to be
in conflict with the rest of the spectrum. A correct
$\rho^\prime-\rho^{\prime\prime}$ splitting induced a far too large
splitting in the $1^{--}$ states of charmonium and bottonium, and visa
versa. Only if the $\rho^\prime$ was regarded to consist of the $n=2$
and $n=3$ states, and the $\rho^{\prime\prime}$ of the $n=4$ and $n=5$
states, correct splittings for the entire spectrum could be obtained. In
addition, the correct splitting between the observed
$\rho^{\prime\prime\prime}$
and $\rho^{iv}$ ($\sim$ 50 MeV) was obtained.
The difference between the $n=2$ and $n=3$ mass, and between the $n=4$
and $n=5$ mass, was found to be $\sim$ 100 Mev, which is much smaller than
the decay widths of the $\rho^\prime$ and the $\rho^{\prime\prime}$ ($\sim$
300 MeV).

The masses of the $\pi$, $K$ and $K^*_0$ are found to be
much to high. As was already
mentioned, this was to be expected, because
the small masses of these particles are believed to be a consequence of
spontaneous breaking of chiral symmetry.
In all cases Ia, Ib and II, the mass of the $\eta_c$ is found to be too
high.
As was pointed out by Hirono et.al. \cite{jappen}, this is problably
a consequence of
neglecting negative energy states.
They found that for a quark model based on the instantaneous ladder BS
equation for charmonium, the $\eta_c$ is strongly influences by neglecting
these states ($\sim$ 100 MeV), while the dependence on all other states
is much weaker ($\sim$ 10 MeV).
If one extrapolates these results to the present theory, this means
that the omission of the negative energy states only weakly affects
the spectrum. Only for the ${}^1S_0$ ground states a substantial mass drop
may arise. This would mean that also the masses of the
$D$ and the $D_S$, which are now found a bit too high, would become
smaller. The $B$ would also get a smaller mass, which only has a positive
result in case Ia.

The centre of gravity COG(n)
(see e.g. Sec.\ 8.1 of \cite{rev}) is defined by:
\begin{displaymath}
{\rm COG}(n)\equiv \frac{5}{9}M(n{}^3P_2)+
\frac{1}{3}M(n{}^3P_1)+\frac{1}{9}M(n{}^3P_0)
\end{displaymath}
It can be proved that, for an arbitrary scalar potential $V_S$ and a
Coulomb vector potential $V_V$, up to first order relativistic
corrections, this COG equals the corresponding $n^1P_1$ singlet. The
relation is violated by the $Q$-dependence of $\alpha_s$ and the presence
of a confining term in the vector direction. It is also
affected by higher order relativistic corrections.
In all cases the COG is found to be somewhat higher than the
corresponding singlet state.
A related quantity is the ratio \cite{rev,rho}
\begin{equation}
\label{rhodef}
\rho=\frac{M({}^3P_2)-M({}^3P_1)}{M({}^3P_1)-M({}^3P_0)}.
\end{equation}
Its experimental value is 0.21 for $u\!/\!d\overline{u\!/\!d}$,
0.48 for $c\overline c$, 0.66 for $n=1$ $b\overline b$ and 0.57 for $n=2$
$b\overline b$. For all three cases Ia, Ib and II, a rather constant
value of $\rho\sim {\rm 0.8}$
(see Table \ref{fitresult}) was found.
A perturbative configuration space calculation shows (see e.g. Sec.\ 4.2 of
\cite{rev}) that this too
large value for $\rho$ is a consequence of the dominance of the vector OGE.
An analysis for the
present case in momentum space, gives a similar result.
A more profound linear scalar potential might lower this ratio.

The following remarks on the parameter sets can be made.
{}From Table\ \ref{fitresult} it is seen that the
quark masses are substantially larger than usual in quark models.
Furthermore, the masses are quite different for the different cases. The
smallest masses are obtained by fit Ia. This is a consequence of
the large negative constant $C\sim -{\rm 1.0}\;{\rm GeV}$, which, however,
is neccesary in order to obtain a good fit for the entire spectrum.
If, for instance one only considers bottonium and charmonium, it turns
out that the quality of the fit only weakly depends on the value of $C$.
The system is overparametrized and one in fact does not even need a
constant in the potential. But, when simultaneously also good results
for the lighter mesons are required, the large negative constant
arises automatically.

The total string tension $\lambda_{\rm tot}$  is defined as the sum of
the tensions in vector and scalar direction. For case I one has
$\lambda_{\rm tot}=\lambda+\frac{1}{2}\alpha_0\Lambda^2$, while case II
simply gives $\lambda_{\rm tot}=\lambda$. These tensions are also quite
different for the different cases. In case Ib, which gave the best fit,
there is only a vector tension. This is in contrast to the requirement
that, in order toobtain
better $\rho$ values, the confining should be dominantly scalar.
The total tension for
case Ia is closest to value $\lambda\sim {\rm 0.18}\;
{\rm GeV}^2$, which is often given in the literature.

The Regge slopes of Ia and Ib are compatible with the experimental value
$\beta\approx {\rm 1.2}\;{\rm GeV}^2$. The slope found in II, is somewhat
too low. This can clearly be seen from the high-$J$ states like the $\rho_5$,
$a_6$ and $K^*_5$. The errors given in Table\
\ref{fitresult} represent a measure of linearity of the trajectories.
It is defined as the spread in the difference between the masses squared
of adjacent states. The spreads found are considerably smaller than the
experimental value.

Finally the running coupling constant for $Q={\rm 31}\;{\rm GeV}$ and
$Q=M_Z={\rm 91.16}\;{\rm GeV}$ were compared with the experimental values
\begin{eqnarray}
&\alpha_s({\rm 34}\;{\rm GeV})&={\rm 0.14}\pm {\rm 0.02},\nonumber\\
&\alpha_s(M_Z)&={\rm 0.1134}\pm {\rm 0.0035}.\nonumber
\end{eqnarray}
Only case Ia is compatible with both conditions.
The choice $\alpha_0={\rm 1.75}$
was made to give the best approximation to the QCD
formula (\ref{alphaqcd}) for moderate momentum transfer. Now it appears
that this choice also gives correct results for very high momenta. A fit
of type Ia, but now with $\alpha_0=16\pi/27$ (not displayed) gave a too
large high momentum $\alpha_s$.
In principle, the high
$Q$-range of the potential is completely irrelevant for the calculation
of the meson spectrum, where the potential is only tested up to a few
GeV. Nevertheless, for the sake of theoretical consistency, this test was
made.

\section{Concluding remarks}
In this paper a relativistic quark model defined in momentum space was
studied.
The quark-antiquark potential used, consisted of a OGE with a
Lorentz vector character, and a linear plus constant confining potential.
For the OGE the Richardson potential $V_R$, given by Eq.\ (\ref{richard}),
with and without its
linear part, as well a
modified Richardson potential $V_M$, defined by Eq.\
(\ref{vh}), was regarded.
Best results were obtained for the Richardson potential
including its linear term (case I).
The linear plus constant potential was given a pure scalar character, i.e.
$\epsilon=0$ in Eq.\ (\ref{vvs}). In
this way, the
confining in the vector direction was induced by the linear
part of $V_R$.
For case I two different fits were made, fit Ia, in which the value of
$\alpha_0$ was fixed to 1.75, and fit Ib, in which $\alpha_0$ was varied,
but the string tension in the scalar direction  $\lambda$ was put equal
to zero.
Also reasonable results were obtained for
$V_{\rm OGE}=V_R$ (case II). Here the
confining potential was given a mixed scalar-vector character.
For the fits Ia, Ib and II, most meson masses, with the exception
of the $\pi$, $K$ and $K^*_0$ were found to be reasonably described by the
model. In case Ia and Ib correct Regge slopes were found, and only in
case Ia a correct strong coupling constant for large momenta was found.
The ratios $\rho$, defined by Eq.\ (\ref{rhodef}), however, were in all
three cases found to be too large.
It is concluded that case Ia
should be preferred.

No detailed comparison with other theories has been made, because the
main purpose of this paper was not so much to improve upon the existing
calculations, but rather to show that results of the same quality could be
obtained using a relativistic theory which is formulated in the momentum
representation.

\acknowledgments

I would like to thank Professor Th. W. Ruijgrok for his important
comments and suggestions.

\appendix
\section{Partial wave decomposition}\label{app:partwave}
In this appendix we give the precise form of the decomposition of a
potential $W$ defined by Eq.\ (\ref{vvs}).
The partial waves,
\begin{displaymath}
V_{ij}^{nJ}=(V_V)^{nJ}_{ij}+(V_S)^{nJ}_{ij},\;\;n=s,t,
\end{displaymath}
defined by Eq.\ (\ref{vdec}) can in general be expressed in terms of the
``spinless'' partial waves $W_V^l$ and $W_S^l$ of $V_V$ and $V_S$
respectively. They are defined by
\begin{equation}\label{vnospin}
W_{V,S}^l(p^\prime,p)=
(2\pi p^\prime p) R(p^\prime,p)
\int_{-1}^{+1}W_{V,S}({\bf p}^\prime,{\bf p})P_l(x)dx,
\end{equation}
with $x=\frac{{\bf p}^\prime\cdot{\bf p}}{p^\prime p}$
and $P_l$ the Legendre polynomial of order $l$.
The quantity $R$ is defined by $R(p^\prime,p)=A_1^\prime A_2^\prime
A_1A_2$, with $A=\sqrt{\frac{E+m}{2E}}$.
If furthermore $b=\frac{p}{E+m}$, then the result is:

\widetext
\subsection{Vectorpotential for the $|s_{1,2}\rangle$
states}
\begin{eqnarray}\label{VVs}
(V_V)^{sJ}_{11}(p^\prime,p) & = &
\left[1+3(b_1^\prime b_2^\prime+b_1b_2)+b^\prime_1b^\prime_2b_1b_2\right]
W_V^J
+(b^\prime_1-b^\prime_2)(b_1-b_2)\frac{(J+1)W_V^{J+1}+JW_V^{J-1}}
{2J+1},\nonumber\\
(V_V)^{sJ}_{12}(p^\prime,p) & = &
(b_2^\prime-b_1^\prime)(b_2+b_1)\sqrt{J(J+1)}\frac{W_V^{J+1}-W_V^{J-1}}
{2J+1}=(V_V)^{sJ}_{21}(p,p^\prime),\\
(V_V)^{sJ}_{22}(p^\prime,p) & = &
(1+b_1^\prime b_2^\prime)(1+b_1b_2)W_V^J+(b_1^\prime+b_2^\prime)(b_1+b_2)
\frac{JW_V^{J+1}+(J+1)W_V^{J-1}}{2J+1}.\nonumber
\end{eqnarray}
\subsection{Vectorpotential for the $|t_{1,2}\rangle$
states}
\begin{eqnarray}\label{VVt}
(V_V)^{tJ}_{11}(p^\prime,p) & = &
(1-b_1^\prime b_2^\prime)(1-b_1b_2)\frac{(J+1)W_V^{J+1}+JW_V^{J-1}}{2J+1}
\nonumber\\&&
+\left[(b_1^\prime-b_2^\prime)(b_1-b_2)+4(b_1^\prime b_2+b_2^\prime b_1)
\right]W_V^J,\nonumber\\
(V_V)^{tJ}_{12}(p^\prime,p) & = &
-(1-b_1^\prime b_2^\prime)(1+b_1b_2)
\sqrt{J(J+1)}\frac{W_V^{J+1}-W_V^{J-1}}{2J+1}=
(V_V)^{tJ}_{21}(p,p^\prime),\\
(V_V)^{tJ}_{22}(p^\prime,p) & = &
(1+b_1^\prime b_2^\prime)(1+b_1b_2)
\frac{JW_V^{J+1}+(J+1)W_V^{J-1}}{2J+1}+(b_1^\prime+b_2^\prime)(b_1+b_2)
W_V^J.\nonumber
\end{eqnarray}
\subsection{Scalarpotential for the $|s_{1,2}\rangle$
states}
\begin{eqnarray}\label{VSs}
(V_S)^{sJ}_{11}(p^\prime,p) & = &
(1+b_1^\prime b_2^\prime b_1b_2)W_S^J-(b_1^\prime b_1+b_2^\prime b_2)
\frac{(J+1)W_S^{J+1}+JW_S^{J-1}}{2J+1},\nonumber\\
(V_S)^{sJ}_{12}(p^\prime,p) & = &
(b_1^\prime b_1-b_2^\prime b_2)
\sqrt{J(J+1)}\frac{W_S^{J+1}-W_S^{J-1}}{2J+1}=
(V_S)^{sJ}_{21}(p,p^\prime),\\
(V_S)^{sJ}_{22}(p^\prime,p) & = &
(1+b_1^\prime b_2^\prime b_1b_2)W_S^J-(b_1^\prime b_1+b_2^\prime b_2)
\frac{JW_S^{J+1}+(J+1)W_S^{J-1}}{2J+1}.\nonumber
\end{eqnarray}
\subsection{Scalarpotential for the $|t_{1,2}\rangle$
states}
\begin{eqnarray}\label{VSt}
(V_S)^{tJ}_{11}(p^\prime,p) & = &
(1+b_1^\prime b_2^\prime b_1b_2)
\frac{(J+1)W_S^{J+1}+JW_S^{J-1}}{2J+1}-(b_1^\prime b_1+b_2^\prime b_2)
W_S^J,\nonumber\\
(V_S)^{tJ}_{12}(p^\prime,p) & = &
-(1-b_1^\prime b_2^\prime b_1b_2)
\sqrt{J(J+1)}\frac{W_S^{J+1}-W_S^{J-1}}{2J+1}=
(V_S)^{tJ}_{21}(p,p^\prime),\\
(V_S)^{tJ}_{22}(p^\prime,p) & = &
(1+b_1^\prime b_2^\prime b_1b_2)
\frac{JW_S^{J+1}+(J+1)W_S^{J-1}}{2J+1}-(b_1^\prime b_1+b_2^\prime b_2)
W_S^J.\nonumber
\end{eqnarray}

\narrowtext
Strictly speaking, these results are only valid for $J>0$. For $J=0$
only the $V_{11}$'s are nonzero and are also given by Eqs.\
(\ref{VVs},..,\ref{VSt}), but with $W^{l=-1}=0$.

In the equal mass case there is no difference between the $b_1$'s and the
$b_2$'s. {}From this it is seen that
the $V^{sJ}_{12}$'s and the $V^{sJ}_{21}$'s are zero.
This means that the potential decouples with regard to the
$|s_1\rangle$ state, which corresponds to the $l=J$ singlet, and the
$|s_2\rangle$ state, which corresponds to the $l=J$ triplet.
Therefore in this case only the $|l=J\pm1\rangle$ triplet states mix. In
the unequal mass case the $l=J$ singlet and triplet states mix.

\section{Spinless decomposition of the
Richardson and Modified Richardson potential}\label{app:rich}
In this appendix the spinless
partial wave decomposition $W_R$ and $W_M$ of the
potentials $V_R$ and $V_M$, defined by
Eqs. (\ref{richard}) and (\ref{vh}) will be calculated.

The momentum transfer $Q^2$ which is defined by Eq.\ (\ref{Qdef})
depends on the lengths
$p=|{\bf p}|$ and $p^\prime=|{\bf p}^\prime|$ of the incoming resp. outgoing
momentum, and on the angle $x=\frac{{\bf p}\cdot {\bf p}^\prime}{pp^\prime}$
between these two momenta:
\begin{equation}\label{qpp}
Q^2=Q^2(p,p^\prime,x)=2pp^\prime(z_0-x).
\end{equation}
Here
\begin{equation}\label{z0}
z_0(p,p^\prime)=\frac{p^2+p^{\prime 2}-\tau(p,p^\prime)}{2pp^\prime},
\end{equation}
where the retardation $\tau$ is a theory dependent quantity which in the
present case is given by Eq.\ (\ref{retard})
The spinless partial wave $W^l(p^\prime,p)$ corresponding to an angular
momentum $l$ is defined by Eq.\ (\ref{vnospin}).
Introducing
\begin{equation}\label{ydef}
y(x)=1+\frac{Q^2(x)}{\Lambda^2},\hspace{1cm}
b=\frac{\Lambda^2}{2pp^\prime}
\end{equation}
and $y_\pm=y(x=\mp 1)\geq 1$, then $W^l_R$ and $W^l_M$ are given by
\begin{eqnarray}\label{vpart}
W_R^l&=&-\frac{\alpha_0R}{2\pi}\int_{y_-}^{y_+}\frac{P_l[z_0-b(y-1)]}
{(y-1)\log y}dy,\\
W_M^l&=&-\frac{\alpha_0R}{2\pi}\int_{y_-}^{y_+}\frac{P_l[z_0-b(y-1)]}
{y\log y}dy.
\end{eqnarray}
The $y$ dependence in $P_l$ can be expanded, using
\begin{equation}\label{gdef}
P_l(z-w)=\sum_{i=0}^lg_i^l(z)w^i.
\end{equation}
Here $g_i^l$ is a polynomial of degree $l-i$. For $l\leq 3$ it is given
in Table I.
For general $l$ it can be found from the recurrence relation
\begin{equation}\label{grec}
(l+1)g_i^{l+1}=(2l+1)(zg_i^l-g_{i-1}^l)-lg_i^{l-1},
\end{equation}
in combination with the initial values
\begin{equation}\label{ginit}
g_0^{-1}(z)=0,\hspace{1cm} g_0^0(z)=1.
\end{equation}
Note that $g^l_0$ obeys the recurrence relation of the Legendre
Polynomials. In combination with
the initial values (\ref{ginit}) it follows
that $g_0^l=P_l$.

The partial waves $W^l_R$ and $W^l_M$
can be written in terms of
$g^l_i$'s and the integrals
\begin{equation}\label{adef}
A_n=\frac{b^n}{2}\int_{y_-}^{y_+}\frac{(y-1)^n}{y\log y}dy,\;
n\geq -1.
\end{equation}
For $n=-1$ one has
\begin{equation}\label{amin1}
A_{-1}=\frac{1}{2b}\left[F(\log y_+)-F(\log y_-)\right],
\end{equation}
where
\begin{equation}\label{fdef}
F(x)\equiv-\int_{x}^\infty\frac{dt}{t(e^t-1)},\;x>0.
\end{equation}
For $n\geq 0$ the integrals $A_n$ can be expanded into
\begin{equation}\label{aexpand}
A_n=\frac{b^n}{2}\sum_{k=0}^n
\left(\begin{array}{c} n\\k\end{array}\right)
(-1)^{(n-k)} I_k,
\end{equation}
where
\begin{equation}\label{idef}
I_n=\int_{y_-}^{y_+}\frac{y^{n-1}}{\log y} dy,\; n\geq 0.
\end{equation}
One finds
\begin{eqnarray}
I_0 & = & \log\left[\frac{\log y_+}{\log y_-}\right],\label{inul}\\
I_n & = & Ei(n\log y_+)-Ei(n\log y_-),\;n>0,\label{in}
\end{eqnarray}
where
\begin{equation}\label{ei}
Ei(x)\equiv -\hspace{-0.4cm}\int_{-\infty}^x\frac{e^t}{t}dt
\end{equation}
is the Exponential integral (see Eq.\ (5.1.2) of \cite{abra}).
The principal value integral is denoted by $-\hspace{-0.3cm}\int$.

Summarizing all steps it follows that $W^l_R$ and $W^l_M$, defined by
Eqs.\ (\ref{vnospin}), (\ref{richard}) and (\ref{vh}) are given by
\begin{eqnarray}
W_R^l(p,p^\prime) & = & -\frac{\alpha_0R}{\pi}
\sum_{i=0}^l g_i^l(z_0)(A_i+bA_{i-1}),\label{vrl}\\
W_M^l(p,p^\prime) & = & -\frac{\alpha_0R}{\pi}
\sum_{i=0}^l g_i^l(z_0)A_i,\label{vhl}
\end{eqnarray}
where the polynomials $g_i^l$ are defined by Eq.\ (\ref{gdef}) and the
integrals $A_n$ can be found from Eqs.\ (\ref{amin1}) and (\ref{aexpand}).

\begin{figure}
\caption{
Running coupling constant $\alpha_s(Q^2)$, defined by
Eq.\ (\protect\ref{alphas})
for three different
choices of $V_{OGE}$ compared to the QCD
formula (\protect\ref{alphaqcd}) and its standard approximation
(\protect\ref{alsapp}).
$\Lambda=\Lambda^{(5)}_{\overline{MS}}={\rm 0.3}\;{\rm GeV}$
and $\alpha_0=16\pi/27$.
}
\label{fig1}
\end{figure}

\narrowtext
\begin{table}
\caption{Regge trajectories calculated from the ultrarelativistic Eq.\
(\protect\ref{ultra}). The masses are expressed in terms of
$\protect\sqrt\lambda$,
where $\lambda$ is the string tension.
}
\begin{tabular}{ddddd}
 &\multicolumn{2}{c}{Present case:}
 &\multicolumn{2}{c}{Basdevant and Boukra:}\\
 &\multicolumn{2}{c}{potential (\protect\ref{wijultra})}
 &\multicolumn{2}{c}{potential (\protect\ref{zijultra})}\\
$l$& $M_l$&$(M^2_{l}-M^2_{l-1})/(8\sqrt 2)$
&$M_l$&$(M^2_{l}-M^2_{l-1})/8$\\
\tableline
0&  3.830 &       & 3.157 &       \\
1&  5.062 & 0.969 & 4.225 & 0.985 \\
2&  6.066 & 0.987 & 5.079 & 0.994 \\
3&  6.931 & 0.993 & 5.811 & 0.996 \\
4&  7.701 & 0.996 & 6.461 & 0.998 \\
5&  8.402 & 0.998 & 7.052 & 0.999 \\
6&  9.049 & 0.998 & 7.597 & 0.998 \\
\end{tabular}
\label{regge}
\end{table}

\begin{table}
\caption{The polynomials $g_i^l$ for $l\leq 3$, defined by Eq.\
(\protect\ref{gdef})}
\begin{tabular}{ccccc}
$l\;\setminus\; i$&0&1&2&3\\
\tableline
0&$1$&&&\\
$1$&$z$&$-1$&&\\
2&$\frac{3}{2}z^2-\frac{1}{2}$&$-3z$&$\frac{3}{2}$&\\
3&$\frac{5}{2}z^3-\frac{3}{2}z$&$-\frac{15}{2}z^2+
\frac{3}{2}$&$\frac{15}{2}z$&
$-\frac{5}{2}$\\
\end{tabular}
\end{table}

\begin{table}
\caption{Final parameter sets from the fitting procedure described in
Sec.\ \protect\ref{sec:num} for potential models I and II.
The varied parameters are indicated by a ``$\bullet$''. For
model I two different fits were made. In case Ia $\alpha_0$
was held fixed and $\lambda$ was fitted,
while in case Ib $\lambda$ was put equal
to $0$ and $\alpha_0$ was fitted.
The related quantities are discussed in Sec.\ \protect\ref{sec:discus}.
}
\label{fitresult}
\begin{tabular}{lddd}
Model:& \multicolumn{1}{c}{Ia}&\multicolumn{1}{c}{Ib}&
\multicolumn{1}{c}{II}\\
\tableline
\underline{Potential}\\
$\alpha_0$	&1.750&2.434$\;\bullet$&         1.862\\
$\Lambda$ (GeV) &0.324$\;\bullet$&0.320$\;\bullet$&0.376$\;\bullet$\\
$\lambda$ (${\rm GeV}^2$)&0.077$\;\bullet$&0      &0.136$\;\bullet$\\
$C$ (GeV)       &-1.297$\;\bullet$&-1.291$\;\bullet$&-1.038$\;\bullet$\\
$\epsilon$      &         0    &         0    &0.523$\;\bullet$\\
\underline{Quark masses}\\
$m_{u\!/\!d}$ (GeV) &0.512$\;\bullet$&0.699$\;\bullet$&0.966$\;\bullet$\\
$m_s$ (GeV)     &0.766$\;\bullet$&0.889$\;\bullet$&1.072$\;\bullet$\\
$m_c$ (GeV)     &2.066$\;\bullet$&2.206$\;\bullet$&2.249$\;\bullet$\\
$m_b$ (GeV)     &5.474$\;\bullet$&5.616$\;\bullet$&5.593$\;\bullet$\\
\tableline
\# parameters     &7&7&8\\
$\chi^2$&263&250&322\\
\underline{Related quantities}\\
$\lambda_{\rm tot}\;({\rm GeV}^2)$&0.169&0.125&1.136\\
$\beta\;({\rm GeV}^2)$&\multicolumn{1}{c}{1.18$\pm$0.05}&
\multicolumn{1}{c}{1.27$\pm$0.09}&
\multicolumn{1}{c}{0.93$\pm$0.08}\\
$\rho$&0.81&0.79&0.75\\
$\alpha_s({\rm 34 GeV})$&0.141&0.196&0.155\\
$\alpha_s(M_Z)$&0.1164&0.161&0.127
\end{tabular}
\end{table}

\pagebreak

\widetext
\begin{table}
\caption{Meson spectrum calculated from Eq.\
(\protect\ref{waveequation}) for three different parameter sets Ia, Ib
and II,
( see Table \protect\ref{fitresult}).
All masses are in (GeV).
The experimental values are taken from
\protect\cite{data}, with the exception of the $h_{c1}$, which is taken
from \protect\cite{hc1}. The mesons labeled with a ``$\bullet$'' (regarded as
beeing established by \protect\cite{data}) were, with the exclusion of
the $D_s^*$ and the $D_{sJ}$, involved in the fitting procedure. The
weights $\sigma_i$ are determined by Eq.\ (\protect\ref{sigma}).
The most dominant ${}^{2s+1}L_J$ waves are underlined.
}
\label{meson_spectrum}
\begin{tabular}{rllldddddd}
\multicolumn{9}{l}{Light unflavoured mesons: $u\!/\!d$ quarks.}\\
\tableline
\multicolumn{2}{c}{Name $i$}& $J^{PC}$ & ${}^{2s+1}L_J$& $M_i^{exp}$ & $n$
&$M_i^{\rm Ia}$&$M_i^{\rm Ib}$&$M_i^{\rm II}$&$\sigma_i$\\
\tableline
$\bullet$&$\pi$
&$0^{-+}$&${}^1S_0$&0.135 &1&
0.600&0.595&0.688&0.400\\
$\bullet$&$\pi^\prime$
&$0^{-+}$&${}^1S_0$&1.300 &2&
1.243&1.206&1.292&0.100 \\
&$\pi^{\prime\prime}$
&$0^{-+}$&${}^1S_0$&1.775 &3&
1.711&1.671&1.695 &\\
\tableline
$\bullet$&$\rho$
&$1^{--}$&${}^3\underline
S_1/{}^3D_1$&0.768&1&
0.754&0.762&0.867&0.020 \\
&&$1^{--}$&${}^3\underline
S_1/{}^3D_1$&      &2&
1.365&1.345&1.387&      \\
$\bullet$&$\rho^\prime$
&$1^{--}$&${}^3S_1/{}^3\underline
D_1$&1.465 &3&
1.474&1.477&1.460&0.025 \\
$\bullet$&$\rho^{\prime\prime}$
&$1^{--}$&${}^3\underline S_1/{}^3D_1$&1.700 &4&
1.806&1.786&1.764&0.020 \\
&&$1^{--}$&${}^3S_1/{}^3\underline
	 D_1$&      &5&
1.865&1.864&1.807&      \\
 &$\rho^{\prime\prime\prime}$
 &$1^{--}$&${}^3\underline S_1/{}^3D_1$&2.100 &6&
2.162&2.151&2.065&      \\
	 &$\rho^{iv}$
	 &$1^{--}$&${}^3S_1/{}^3\underline
	 D_1$&2.150 &7&
2.200&2.206&2.096&      \\
\tableline
$\bullet$&$a_0$
&$0^{++}$&${}^3P_0$&0.983&1&
1.012&0.981&1.017&0.020 \\
	 &$a_0^\prime$
	 &$0^{++}$&${}^3P_0$&1.320 &2&
1.517&1.464&1.510&      \\
$\bullet$&$a_1$
&$1^{++}$&${}^3P_1$&1.260 &1&
1.166&1.163&1.197&0.030 \\
$\bullet$&$a_2$
&$2^{++}$&${}^3\underline P_2/{}^3F_2$&1.318 &1&
1.301&1.319&1.329&0.020 \\
\tableline
 &COG	&        & ${}^3P_{0,1,2}$&1.262 &1&
1.224&1.229&1.250&     \\
$\bullet$&$b_1$
&$1^{+-}$&${}^1P_1$&1.232 &1&
1.183&1.194&1.231&0.020 \\
\tableline
$\bullet$&$\pi_2$			&$2^{-+}$&${}^1D_2$&1.670 &1&
1.590&1.614&1.561&0.020 \\
         &$\pi_2^\prime$		&$2^{-+}$&${}^1D_2$&2.100 &2&
1.958&1.986&1.880&      \\
\tableline
$\bullet$&$\rho_3$&$3^{--}$&${}^3\underline D_3/{}^3G_3$&1.691 &1&
1.698&1.734&1.637&0.020 \\
         &$\rho_3^\prime$&$3^{--}$&${}^3\underline D_3/{}^3G_3$&2.250 &2&
2.051&2.092&1.940&      \\
	 &				&$3^{--}$&${}^3D_3/{}^3\underline
	 G_3$&      &3&
2.097&2.152&1.969&      \\
\tableline
         &$a_3$				&$3^{++}$&${}^3F_3$&2.050 &1&
1.915&1.957&1.813&      \\
         &$a_4$		&$4^{++}$&${}^3\underline F_4/{}^3H_4$&2.040 &1&
2.021&2.085&1.883&      \\
	 &$\rho_5$	&$5^{--}$&${}^3\underline G_5/{}^3I_5$&2.350 &1&
2.297&2.395&2.093&      \\
	 &$a_6$		&$6^{++}$&${}^3\underline H_6/{}^3J_6$&2.450 &1&
2.540&2.677&2.279&\\
\tableline\\
\multicolumn{9}{l}{Strange mesons (Kaons): $s$, $u\!/\!d$
quarks.}\\
\tableline
\multicolumn{2}{c}{Name $i$}& $J^{PC}$ & ${}^{2s+1}L_J$& $M_i^{exp}$ & $n$
&$M_i^{\rm Ia}$&$M_i^{\rm Ib}$&$M_i^{\rm II}$&$\sigma_i$\\
\tableline
$\bullet$&$K$				&$0^-$&${}^1S_0$&0.495 &1&
0.762&0.723&0.781&0.200\\
	 &$K^\prime$			&$0^-$&${}^1S_0$&1.460 &2&
1.402&1.329&1.385&     \\
	 &$K^{\prime\prime}$		&$0^-$&${}^1S_0$&1.830 &3&
1.864&1.786&1.786&\\
\tableline
$\bullet$&$K^*$		&$1^-$&${}^3\underline S_1/{}^3D_1$&0.894 &1&
0.891&0.876&0.955&0.020\\
$\bullet$&$K^{*\prime}$	&$1^-$&${}^3\underline S_1/{}^3D_1$&1.412 &2&
1.504&1.455&1.477&0.020\\
$\bullet$&$K^{*\prime\prime}$&$1^-$&${}^3S_1/{}^3\underline D_1$&1.714 &3&
1.618&1.588&1.553&0.020\\
	 &	&$1^-$&${}^3\underline S_1/{}^3D_1$&      &4&
1.945&1.890&1.852&\\
\tableline
$\bullet$&$K_0^*$			&$0^+$&${}^3P_0$&1.429 &1&
1.177&1.112&1.115&0.200\\
	 &				&$0^+$&${}^3P_0$&      &2&
1.674&1.587&1.604&     \\
	 &$K_0^{*\prime}$		&$0^+$&${}^3P_0$&1.950 &3&
2.074&1.989&1.955&\\
\tableline
$\bullet$&$K_1$	&$1^+$&${}^1\underline P_1/{}^3P_1$&1.270 &1&
1.304&1.274&1.288&0.020\\
$\bullet$&$K_1^\prime$&$1^+$&${}^1P_1/{}^3\underline P_1$&1.402 &2&
1.322&1.306&1.321&0.020\\
 &$K_1^{\prime\prime}$	&$1^+$&${}^1\underline P_1/{}^3P_1$&1.650 &3&
1.773&1.725&1.701&     \\
\tableline
$\bullet$&$K_2^*$&$2^+$&${}^3\underline P_2/{}^3F_2$&1.429 &1&
1.416&1.415&1.415&0.020\\
	 &$K_2^{*\prime}$&$2^+$&${}^3\underline P_2/{}^3F_2$&1.980 &2&
1.867&1.849&1.785&     \\
	 &		&$2^+$&${}^3P_2/{}^3\underline F_2$&      &3&
1.945&1.934&1.831&     \\
\tableline
	 &$K_2$		&$2^-$&${}^1\underline D_2/{}^3D_2$&1.580 &1&
1.706&1.693&1.640&     \\
$\bullet$&$K_2^\prime$	&$2^-$&${}^1D_2/{}^3\underline D_2$&1.768 &2&
1.715&1.711&1.650&0.020\\
 &$K_2^{\prime\prime}$	&$2^-$&${}^1\underline D_2/{}^3D_2$&2.250 &3&
2.082&2.065&1.962&     \\
\tableline
$\bullet$&$K_3^*$			&$3^-$&${}^3D_3/{}^3G_3$&1.770 &1&
1.801&1.813&1.722&0.020\\
 &$K_3$		&$3^+$&${}^1\underline F_3/{}^3F_3$&2.320 &1&
2.027&2.034&1.900&     \\
$\bullet$&$K_4^*$&$4^+$&${}^3\underline F_4/{}^3H_4$&2.045 &1&
2.115&2.150&1.967&0.020\\
	 &$K_4$	&$4^-$&${}^1G_4/{}^3\underline G_4$&2.500 &1&
2.300&2.333&2.116&     \\
	 &$K_5^*$&$5^-$&${}^3G_5/{}^3\underline I_5$&2.380 &1&
2.386&2.449&2.176&     \\
\tableline\\
\multicolumn{9}{l}{Charmed mesons: $c$, $u\!/\!d$ quarks.}\\
\tableline
\multicolumn{2}{c}{Name $i$}& $J^{PC}$ & ${}^{2s+1}L_J$& $M_i^{exp}$ & $n$
&$M_i^{\rm Ia}$&$M_i^{\rm Ib}$&$M_i^{\rm II}$&$\sigma_i$\\
\tableline
$\bullet$&$D$		&$0^-$&${}^1S_0$&1.867&1&
1.935&1.901&1.904&0.020\\
$\bullet$&$D^*$		&$1^-$&${}^3\underline S_1/{}^3D_1$&2.010&1&
2.006&1.999&2.031&0.020\\
$\bullet$&$D_1$		&$1^+$&${}^1\underline P_1/{}^3P_1$&2.424&1&
2.406&2.379&2.382&0.020\\
     &$D_J$(?)	&$1^+$&${}^1P_1/{}^3\underline P_1$&2.440&2&
2.439&2.438&2.424&\\
$\bullet$&$D_2^*$	&$2^+$&${}^3\underline P_2/{}^3F_2$&2.459&1&
2.485&2.492&2.484&0.020\\
\tableline\\
\multicolumn{9}{l}{Charmed strange mesons: $c$, $s$ quarks.}\\
\tableline
\multicolumn{2}{c}{Name $i$}& $J^{PC}$ & ${}^{2s+1}L_J$& $M_i^{exp}$ & $n$
&$M_i^{\rm Ia}$&$M_i^{\rm Ib}$&$M_i^{\rm II}$&$\sigma_i$\\
\tableline
$\bullet$&$D_s$		&$0^-$&${}^1S_0$&1.969&1&
2.032&1.990&1.984&0.020\\
$\bullet$&$D_s^*$(?)	&$1^-$&${}^3\underline S_1/{}^3D_1$&2.110&1&
2.100&2.088&2.110&\\
$\bullet$&$D_{s1}$	&$1^+$&${}^1\underline P_1/{}^3P_1$&2.537&1&
2.498&2.473&2.466&0.020\\
	 &		&$1^+$&${}^1P_1/{}^3\underline P_1$&     &2&
2.520&2.516&2.503&\\
$\bullet$&$D_{sJ}$(?)	&$2^+$&${}^3\underline P_2/{}^3F_2$&2.564&1&
2.561&2.568&2.563&\\
\tableline\\
\multicolumn{9}{l}{Bottom mesons: $b$, $u\!/\!d$ quarks.}\\
\tableline
\multicolumn{2}{c}{Name $i$}& $J^{PC}$ & ${}^{2s+1}L_J$& $M_i^{exp}$ & $n$
&$M_i^{\rm Ia}$&$M_i^{\rm Ib}$&$M_i^{\rm II}$&$\sigma_i$\\
\tableline
$\bullet$&$B$		&$0^-$&${}^1S_0$&5.279 &1&
5.303&5.268&5.247&0.020\\
$\bullet$&$B^*$		&$1^-$&${}^3\underline S_1/{}^3D_1$&5.325 &1&
5.336&5.318&5.316&0.020\\
\tableline\\
\multicolumn{9}{l}{Charmonium: $c$ quarks.}\\
\tableline
\multicolumn{2}{c}{Name $i$}& $J^{PC}$ & ${}^{2s+1}L_J$& $M_i^{exp}$ & $n$
&$M_i^{\rm Ia}$&$M_i^{\rm Ib}$&$M_i^{\rm II}$&$\sigma_i$\\
\tableline
$\bullet$&$\eta_c$			&$0^{-+}$&${}^1S_0$&2.979 &1&
3.042&3.010&3.007&0.020\\
	 &$\eta_c^\prime$		&$0^{-+}$&${}^1S_0$&3.590 &2&
3.615&3.589&3.609&     \\
\tableline
$\bullet$&$J/\Psi$	&$1^{--}$&${}^3\underline S_1/{}^3D_1$&3.097 &1&
3.099&3.104&3.117&0.020\\
$\bullet$&$\Psi^\prime$	&$1^{--}$&${}^3\underline S_1/{}^3D_1$&3.686 &2&
3.655&3.646&3.665&0.020\\
$\bullet$&$\Psi^{\prime\prime}$		&$1^{--}$&${}^3S_1/{}^3\underline
D_1$&3.770 &3&
3.766&3.780&3.775&0.020\\
$\bullet$&$\Psi^{\prime\prime\prime}$
&$1^{--}$&${}^3\underline S_1/{}^3D_1$&4.040 &4&
4.051&4.017&4.028&0.020\\
$\bullet$&$\Psi^{iv}$			&$1^{--}$&${}^3S_1/{}^3\underline
D_1$&4.159 &5&
4.124&4.105&4.097&0.020\\
$\bullet$&$\Psi^{v}$	&$1^{--}$&${}^3\underline S_1/{}^3D_1$&3.415 &6&
4.376&4.319&4.314&0.020\\
	 &		&$1^{--}$&${}^3S_1/{}^3\underline
	 D_1$&      &7&
4.430&4.384&4.364&     \\
\tableline
$\bullet$&$\chi_{c0}$			&$0^{++}$&${}^3P_0$&3.415 &1&
3.437&3.433&3.409&0.020\\
$\bullet$&$\chi_{c1}$			&$1^{++}$&${}^3P_1$&3.511 &1&
3.485&3.506&3.504&0.020\\
$\bullet$&$\chi_{c2}$	&$2^{++}$&${}^3\underline P_2/{}^3F_2$&3.556 &1&
3.523&3.562&3.572&0.020\\
\tableline
	 &COG		&        &${}^3P_{0,1,2}$&3.525 &1&
3.501&3.529&3.531&     \\
	 &$h_{c1}$		&$1^{+-}$&${}^1P_1$&3.526 &1&
3.492&3.520&3.522&\\
\tableline\\
\multicolumn{9}{l}{Bottonium: $b$ quarks.}\\
\tableline
\multicolumn{2}{c}{Name $i$}& $J^{PC}$ & ${}^{2s+1}L_J$& $M_i^{exp}$ & $n$
&$M_i^{\rm Ia}$&$M_i^{\rm Ib}$&$M_i^{\rm II}$&$\sigma_i$\\
\tableline
$\bullet$&$\Upsilon$	&$1^{--}$&${}^3\underline S_1/{}^3D_1$&9.460 &1&
9.493&9.434&9.441&0.010\\
$\bullet$&$\Upsilon^\prime$
&$1^{--}$&${}^3\underline S_1/{}^3D_1$&10.023 &2&
10.011&10.018&10.022&0.010\\
 &		&$1^{--}$&${}^3S_1/{}^3\underline
	 D_1$&       &3&
10.131&10.171&10.160&     \\
$\bullet$&$\Upsilon^{\prime\prime}$
&$1^{--}$&${}^3\underline S_1/{}^3D_1$&10.355 &4&
10.346&10.348&10.365&0.010\\
 &		&$1^{--}$&${}^3S_1/{}^3\underline
	 D_1$&       &5&
10.423&10.444&10.451&     \\
$\bullet$&$\Upsilon^{\prime\prime\prime}$&
$1^{--}$&${}^3\underline S_1/{}^3D_1$&10.580&6&
10.614&10.599&10.626&0.010\\
 &			&$1^{--}$&${}^3S_1/{}^3\underline
	 D_1$&       &7&
10.672&10.670&10.688&     \\
$\bullet$&$\Upsilon^{iv}$&
$1^{--}$&${}^3\underline S_1/{}^3D_1$&10.865 &8&
10.846&10.811&10.844&0.010\\
	 &				&$1^{--}$&${}^3S_1/{}^3\underline
	 D_1$&       &9&
10.893&10.868&10.892&     \\
$\bullet$&$\Upsilon^{v}$&$1^{--}$&${}^3\underline S_1/{}^3D_1$&11.019 &10&
11.054&11.000&11.035&0.010\\
\tableline
$\bullet$&$\chi_{b0}$			&$0^{++}$&${}^3P_0$&9.860  &1&
9.859&9.863&9.843&0.010\\
$\bullet$&$\chi_{b0}$			&$0^{++}$&${}^3P_0$&10.232 &2&
10.220&10.229&10.232&0.010\\
\tableline
$\bullet$&$\chi_{b1}$			&$1^{++}$&${}^3P_1$& 9.892 &1&
9.882&9.906&9.888&0.010\\
$\bullet$&$\chi_{b1}$			&$1^{++}$&${}^3P_1$&10.255 &2&
10.239&10.258&10.261&0.010\\
\tableline
$\bullet$&$\chi_{b2}$&$2^{++}$&${}^3\underline P_2/{}^3F_2$& 9.913 &1&
9.901&9.938&9.922&0.010\\
$\bullet$&$\chi_{b2}$&$2^{++}$&${}^3\underline P_2/{}^3F_2$&10.268 &2&
10.253&10.281&10.284&0.010\\
\tableline
	 &COG		&        &${}^3P_{0,1,2}$& 9.900 &1&
9.890&9.919&9.902&     \\
	 &COG		&        &${}^3P_{0,1,2}$&10.260 &2&
10.245&10.267&10.271&     \\
\end{tabular}
\end{table}

\end{document}